# THE VORTICITY OF SOLAR PHOTOSPHERIC FLOWS ON THE SCALE OF GRANULATION.

A.A. Pevtsov[1]

[1]National Solar Observatory, Sunspot, NM UNITED STATES

e-mail apevtsov@nso.edu

We employ time sequences of images observed with a G-band filter ($\lambda$4305Å) by the Solar Optical Telescope (SOT) on board of Hinode spacecraft at different latitude along solar central meridian to study vorticity of granular flows in quiet Sun areas during deep minimum of solar activity. Using a feature correlation tracking (FCT) technique, we calculate the vorticity of granular-scale flows. Assuming the known pattern of vertical flows (upward in granules and downward in intergranular lanes), we infer the sign of kinetic helicity of these flows. We show that the kinetic helicity of granular flows and intergranular vortices exhibits a weak hemispheric preference, which is in agreement with the action of the Coriolis force. This slight hemispheric sign asymmetry, however, is not statistically significant given large scatter in the average vorticity. The sign of the current helicity density of network magnetic fields computed using full disk vector magnetograms from the Synoptic Optical Long-term Investigations of the Sun (SOLIS) does not show any hemispheric preference. The combination of these two findings suggests that the photospheric dynamo operating on the scale of granular flows is non-helical in nature.

1. INTRODUCTION

The magnetic field on the Sun and sun-like stars is the product of processes collectively called a dynamo. Different dynamos can take place in flow dominated (e.g., solar/stellar convection zone) or magnetic field dominated environments (solar/stellar coronae, Blackman and Ji, 2006). On the Sun, a flow dominated (kinematic) dynamo can operate at the interface of the convection and radiation zones, as well as in the bulk of the convection zone. A separate dynamo may also take place at the interface of the convection zone and photosphere or in the photosphere itself. Observations sup-port the existence of two types of dynamos on the Sun. For example, the strong magnetic fields of active regions have a lifetime of weeks or months, and these fields show strong variations in area and field strength through the solar cycle. On the other hand, quiet Sun fields are much weaker; they have very short lifetimes of minutes and hours, and show very little variation with the solar cycle [Pevtsov and Acton, 2001].

Dynamos can also be classified by their helicity content. [Pevtsov and Longcope, 2007] have provided arguments based on Rossby number that the dynamo operating in lower part of the convection zone should be helical in nature, while a (near) surface dynamo should be non-helical. The latter has been referred in the literature as "local", "surface", "photospheric", "turbulent" or "chaotic" dynamo.

Numerical simulations indicate that a helical dynamo is much more efficient in producing strong magnetic fields [Cattaneo, 1999]. Turbulent dynamos can only amplify the magnetic field to 10-20% of the kinetic energy of dynamo flows [Cattaneo, 1999; Schekochihin et al., 2004]. The typical magnetic energy of the network field at the granular scale is about $10^{27}$ erg, and the kinetic energy of the granular flows is ~3-6 x $10^{27}$ erg. Therefore, granular motions do have sufficient energy to amplify network-type magnetic fields. On the other hand, the magnetic energy of a typical active region is about 1-10 x$10^{34}$ erg. Fields of that strength require helical dynamo as their origin.

[Pevtsov and Longcope, 2001] have proposed an observational test for the two types of dynamos operating on the Sun. They have argued that with sufficient spatial averaging, chaotic dynamos should not exhibit hemispheric preference in the kinetic helicity of dynamo flows (granular scale), and therefore, magnetic fields generated by a (near) surface dynamo should not follow the hemispheric helicity rule discovered in active region magnetic fields [Pevtsov et al., 1995; Abramenko et al., 1996]. Indeed, a visual analysis of high resolution movies of solar granulation suggests a lack of hemisphere-dependent patterns in vorticity. Fig. 1, taken from a time sequence of G-band images made with the Dunn Solar Telescope at Sacramento Peak, shows that bright features found in intergranular lanes may exhibit both clockwise (CW) and counterclockwise (CCW) rota-tion. A time sequence of images of this area (see accompanying movie at http://dx.doi.org/10.7910/DVN/K4XPKK) indicates a complex rotation pattern, with some features reversing the sense of their rotation from CW to CCW.

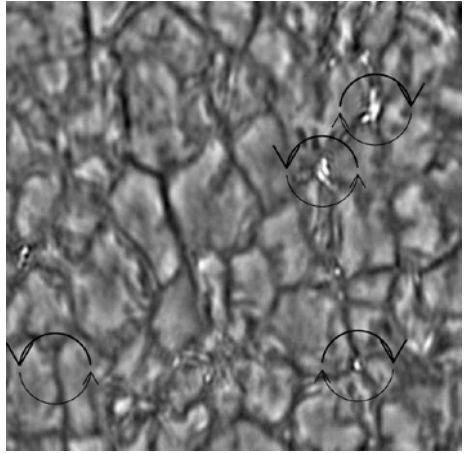

**Рис. 1.** Snapshot of solar granulation taken with the Dunn Solar Telescope's (DST) G-band filter. Ringed features are G-band bright points usually associated with magnetic field concentrations in the photosphere. Black arcs are drawn around selected bright points indicating the sense of the rotational motions. Bright point at the center of the rotational circle shown in low-left corner of the image appears in later images of this sequence.

Using vector magnetograms from the Advanced Stokes Polarimeter at the National Solar Observatory at Sacramento Peak, [Pevtsov and Longcope, 2001] have shown that network magnetic fields exhibit a weak hemispheric preference in their twist (current helicity). This twist was interpreted as the effect of dissipating active regions whose magnetic fields contribute strongly to the network fields. [Pevtsov and Longcope, 2001] have concluded that a chaotic dynamo simply "recycles" magnetic field created by the subphotospheric helical dynamo. This previous study, however, did not measure the kinetic helicity of granular flows directly.

In this paper, we present direct measurements of vorticity and kinetic helicity of granular flows. In Section 2 we describe our data. Sections 3-4 are devoted to calculations of kinetic and current helicities, and Section 5 discusses our findings.

2. **OBSERVATIONS**

Our primary data sets consist of time sequences from the spaceborne Solar Optical Telescope (SOT) on board the Hinode spacecraft. The data were taken on 25-26 November 2008 using the broadband filter imager (BFI) with its G-band filter (λ4305Å). Data were taken at five locations centered at different latitudes along the Sun's central meridian: W00S50, W00S30, W00N00, W00N30, and W00N50, where W, S, and N denote West, South, and North followed by solar central meridian distance and latitude in solar degrees. At each location, we took 50 images with a time cadence of 30 seconds, each image is about 50 arc sec by 200 arc sec in size and with spatial resolution of 0.2 arc sec per pixel.

Images taken in G-band show photospheric layers of the Sun; for that reason, in the following discussion we refer to both as "white-light" observations. The G-band mean height of formation (~54 km above continuum) also results in higher contrast (as compared to the continuum) in granulation pattern. In addition, G-band radiation is enhanced in G-band bright points, which are usually associated with concentrations of magnetic flux.

The observations were taken during the extended period of deep solar minimum with no active region magnetic fields present on the disk (see Fig. 2).

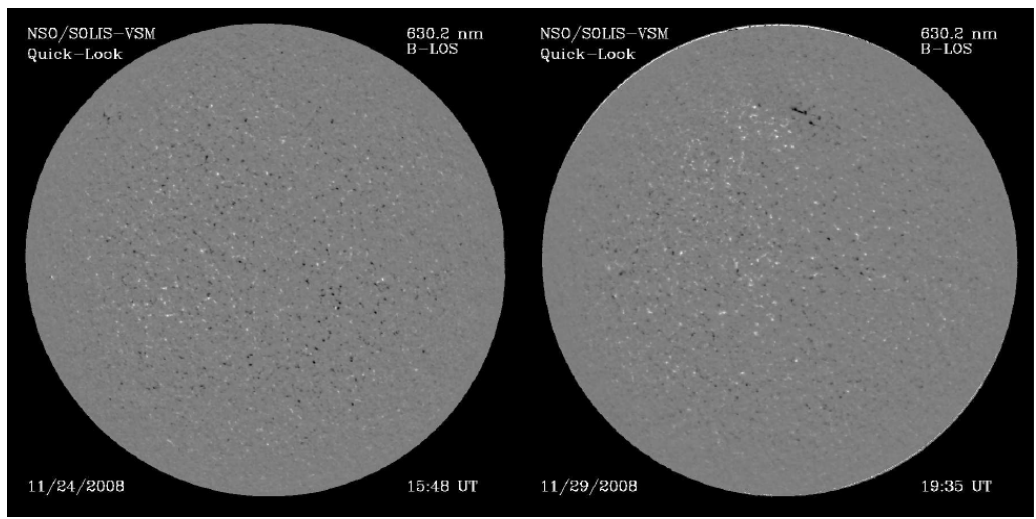

**Рис. 2.** Longitudinal photospheric magnetograms from VSM/SOLIS taken on (left) 24 November 2008 and (right) 29 November 2008. Magnetic fields (white is positive polarity and black is negative polarity) are scaled between +/- 100 Gauss to demonstrate that no strong magnetic fields (of dissipating active regions) are present during Hinode observations.

In addition to white-light observations, we use full disk vector magnetograms from the Vector Spectromagnetograph (VSM), one of three instruments comprising the Synoptic Optical Long-term Investigations of the Sun (SOLIS) synoptic facility at NSO [Keller et al., 1998]. Magnetograms were taken in the photospheric spectral line Fe I λ6302Å, with pixel size 1.13 x 1.13 arc seconds.

### 3. VORTICITY OF GRANULAR FLOWS

Calculation of the kinetic helicity, $h_k = (\nabla \times \mathbf{V}) \cdot \mathbf{V} = \omega \cdot \mathbf{V}$ requires knowledge of the velocity vector, $\mathbf{V}$ of fluid motions and their derivatives in all three directions. Observations in a single layer of the solar atmosphere can be used to derive the vorticity of horizontal flows, $\omega_z = \frac{\partial V_y}{\partial x} - \frac{\partial V_x}{\partial y}$, where the $V_{x,y}$ - components of the horizontal velocity can be derived using a correlation tracking technique. Although correlation tracking does not have information on vertical velocities, one can make a reasonable assumption based on known physics of granular flows that the centers of granules should predominantly be associated with up-flows of material, while intergranular lanes are locations of down-flows. Combining vorticity with the sign of vertical flows allows one to derive the sign of the kinetic helicity associated with granular motions.

To calculate the vorticity, we employ Fourier local correlation tracking (FLCT, Fisher and Welsch, 2008) to track horizontal displacements of granular patterns between successive images. We used a Gaussian apodization window with a width of 10 pixels to smooth images prior to computing the cross-correlations and deriving displacements between images. Since its introduction, the method has been extensively tested and used by several researchers to track horizontal motions of various features in the solar photosphere. Using horizontal displacements, we have calculated vorticity maps for each observed location on the Sun [Rieutord et al., 2001] had compared the model velocity field from their numerical simulations to ones derived from the same data using a correlation tracking technique. They have concluded that at small time (< 0.5 hours) and spatial (< 2.5 Mm) scales granules do not represent large-scale (meso-/supergranular size) flows very well. On smaller scales, the effects of interaction between granules and granular proper motions make the large-scale flows untraceable. However, in our study, we are interested in small-scale flows at the boundary of granules and intergranular lanes. The large-scale flows are ignored because neither our temporal nor spatial cadence is sufficient to trace them. Fig. 3 gives example of solar granulation observed by SOT/Hinode and corresponding vorticity patterns derived from these data. It is clear that patches of both negative and positive vorticity are present. The intergranular lanes show a much stronger vorticity although the data do indicate the presence of weak helicity at the center of granules. A detailed comparison of granulation and vorticity patterns implies that vorticity of both negative and positive sign can be equally present in intergranular lanes (for illustration, compare areas inside the black circles in Fig. 3).

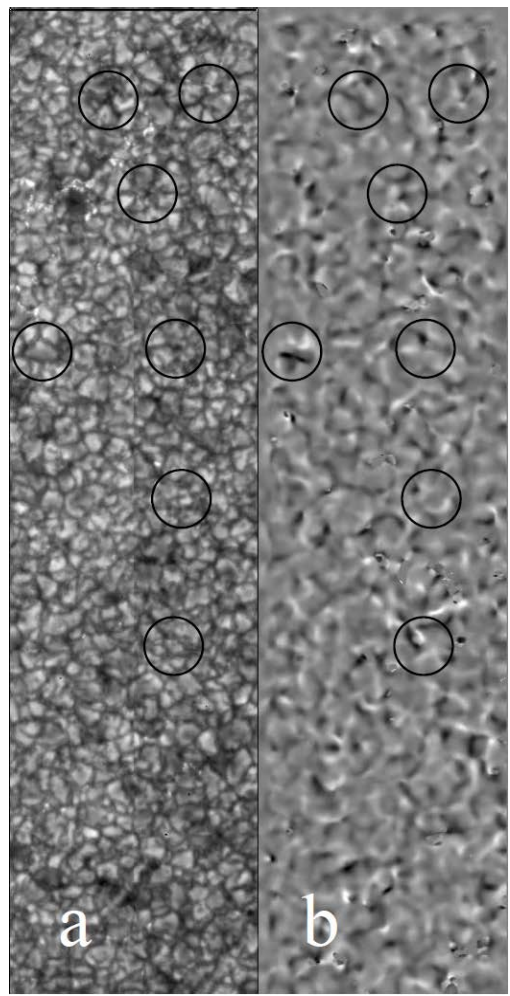

**Рис. 3.** Hinode broadband image of granulation (a) and corresponding pattern of kinetic helicity (b). White (black) halftones indicate positive (negative) helicity. Black circles identify selected examples to compare location of helicity patterns with respect to granular flows.

In the search for asymmetry in the vorticities between the centers and periphery of granules, we have calculated an average intensity of broadband images. We use this average intensity as a discriminator between the center of granules (pixels with intensity above mean intensity) and the intergranular lanes (brightness below the mean intensity). The averaged vorticity computed separately for the center and periphery of granules does show a weak latitudinal variation (Fig. 4). This hemispheric tendency is opposite for the center of granules and the intergranular lanes. Assuming upward flows in granules and downward flows in the intergranular lanes, Fig. 4 suggests the presence of a weak hemispheric dependency of kinetic helicity ($h_k$) of granular flows with positive (negative) $h_k$ in southern (northern) hemisphere. The sign preference is in agreement with the expected action of the Coriolis force. This interesting tendency, however, is very weak and is not statistically significant when one takes into account very large scatter in average vorticity (about 100 times larger than mean values for $h_k$ shown on Fig. 4.

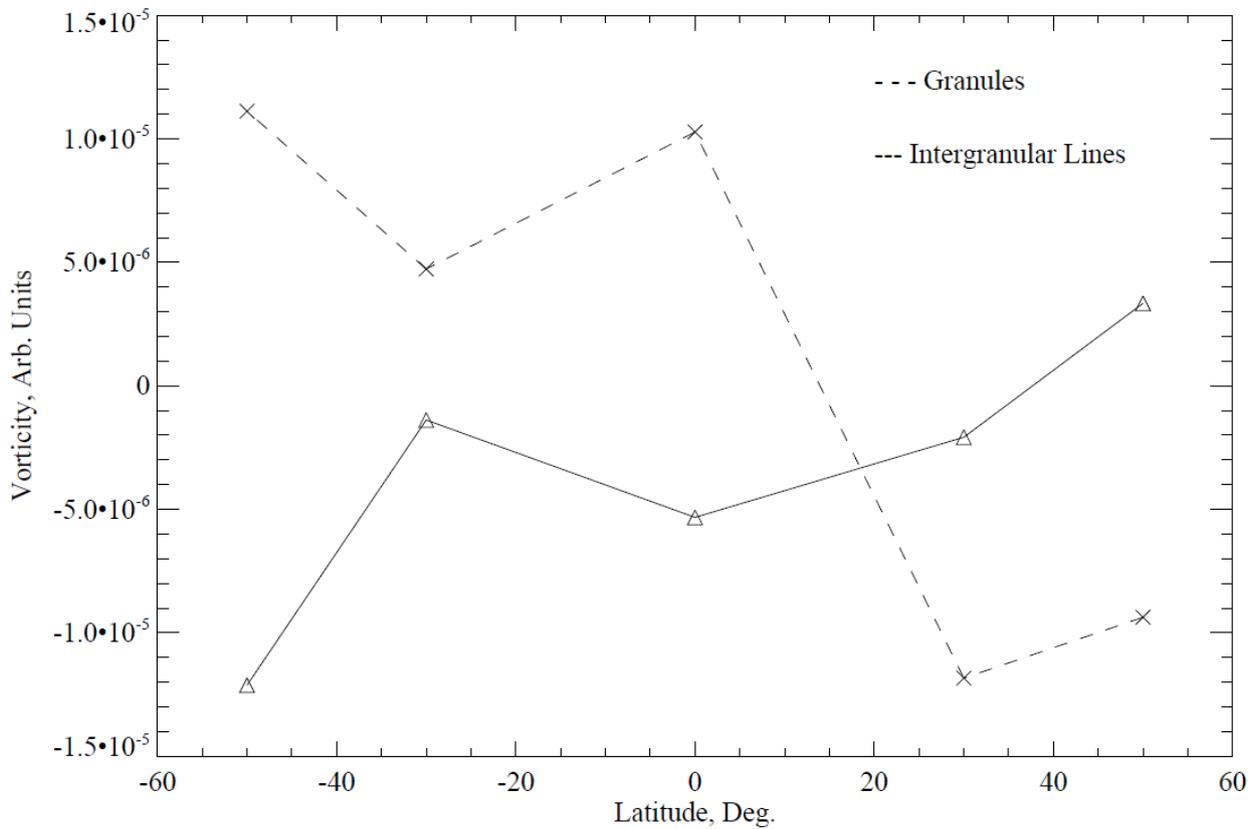

**Рис. 4**. Latitudinal profile of the averaged vorticity for granules (crosses) and intergranular lanes (triangles) on 25-26 November 2008. Vorticity is shown in arbitrary units.

## 4. LATITUDINAL PROFILE OF THE CURRENT HELICITY DENSITY

To compute the current helicity of magnetic fields, we employ full disk vector magnetograms from SOLIS. We use observations from 29 November 2008 because there were no data for 25-26 November. There were no strong fields on the Sun during 25-28 November, and the overall appearance of the longitudinal magnetograms was very similar between 25-26 November and 29 November. To characterize the current helicity, we use the helicity proxy $J_z \cdot B_z$ [Pevtsov et al., 1994], where $B_z$, $J_z$ are the vertical components of the magnetic flux and electric current density. $J_z \cdot B_z$ was computed for each pixel in the SOLIS magnetograms and was averaged over 5 x 5 degrees in solar latitude and longitude. Fig. 5a shows the latitudinal profile of the current helicity density along the solar central meridian. For comparison, we also show the averaged profile of current helicity density along the solar equator (Fig. 5b). In both cases, we see no asymmetry in the sign of current helicity either between northern and southern hemispheres or between eastern and western hemispheres. The lack of northern/southern hemispheric asymmetry in the current helicity is in agreement with the (lack of) hemispheric asymmetric in the kinetic helicity of granular flows (Fig. 4).

## 5. DISCUSSION

The results presented in the previous sections indicate the absence of the hemispheric helicity rule in network magnetic fields. Strong magnetic fields of active regions obey such a rule albeit with a significant scatter [Pevtsovel at., 1995]. Plasma motions associated with granulation-scale flows show a very weak hemispheric sign-asymmetry in their kinetic helicity. This asymmetry, however, is statistically insignificant given the amplitude of standard deviation of mean helicity. Weak or nearly absent hemispheric dependency of kinetic helicity is in agreement with arguments presented by [Pevtsov and Longcope, 2001; 2007]. Taken together, latitudinal profiles of kinetic helicity and current helicity proxy $\alpha_z$ indicate that the photospheric dynamo is non-helical (chaotic) in its nature. A previous investigation by [Pevtsov and Longcope, 2001] had found a weak hemispheric asymmetry in sign of $\alpha_z$ which was in agreement with the hemispheric helicity rule for active regions. Although [Pevtsov and Longcope, 2001] took observations in quiet Sun areas outside of active regions, sunspot activity was high in 2000, and therefore, dissipating fields of active regions could affect significantly the helical properties of the network fields. Observations presented in this paper were taken during deepest part of the current solar minimum, when no sunspot activity was present for extended periods of time. Therefore, the helical properties of network magnetic field during this period reflect action of the photospheric dynamo alone. This dynamo appears to be non-helical in nature.

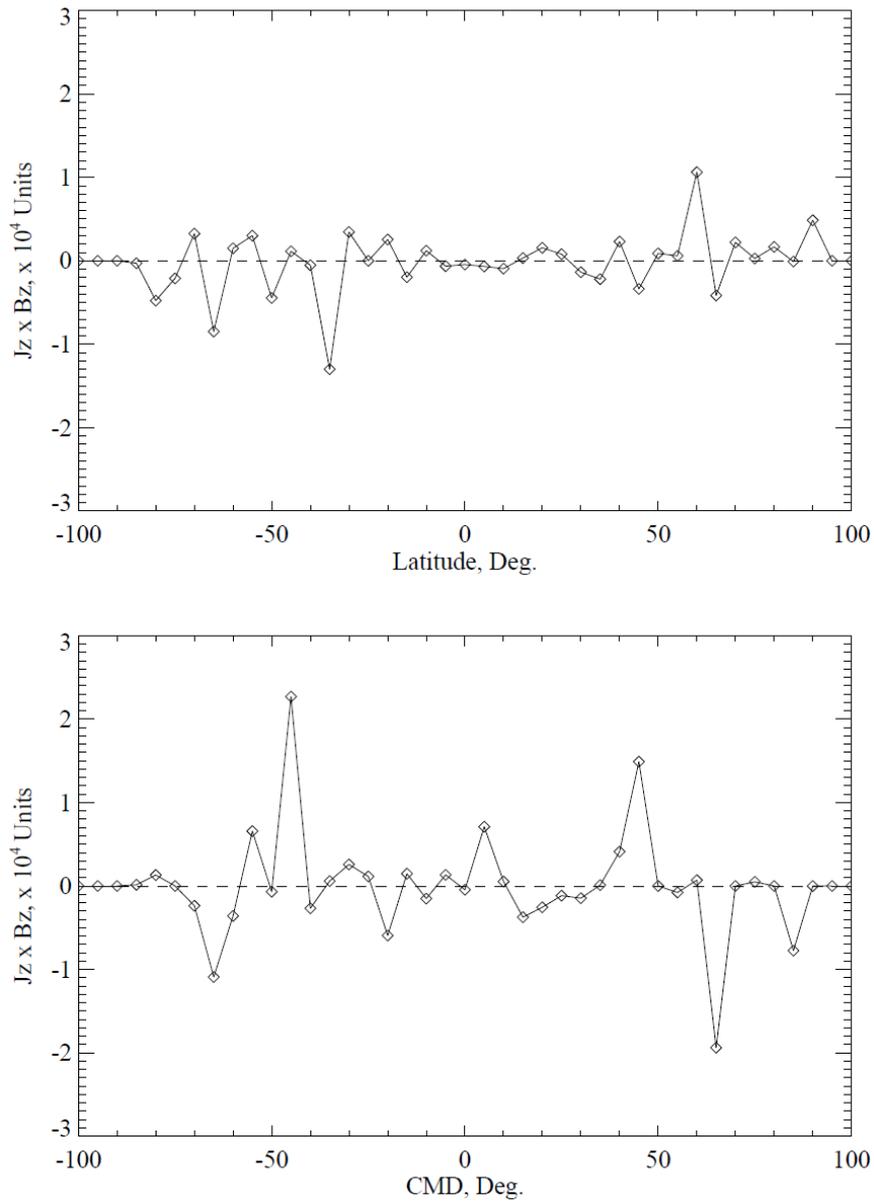

**Рис. 5.** Latitudinal (top panel) and longitudinal (bottom panel) profiles of the averaged current helicity 29 November 2008 from SOLIS vector magnetogram. Current helicity is shown in arbitrary units.